\documentclass[
nofootinbib,
 amsmath,amssymb,
 aps,
]{revtex4-2}
\usepackage{graphicx} % Required for inserting images
\usepackage{dcolumn}% Align table columns on decimal point
\usepackage{bm}
\usepackage{xcolor}

\usepackage[colorlinks,urlcolor=blue,citecolor=blue]{hyperref}

\begin{document}

\title{Non-BPS Monopoles and Dyons via Resurgent Transseries}
\author{Gerald V. Dunne}
\author{Evan Shinn}
\affiliation{Department of Physics, University of Connecticut, Storrs, CT 06269}
%\date{}

\begin{abstract}
 Radially symmetric non-BPS 't Hooft-Polyakov monopoles and dyons are constructed as resurgent transseries: infinite sums of exponentially decaying terms, each multiplied by a factorially divergent fluctuation factor. All higher exponential terms are explicitly expressed in terms of the leading order solutions. In the BPS limit all fluctuation terms truncate. 
    
\end{abstract}

\maketitle

\section{Introduction}

Semiclassical analysis of quantum field theory is a well-developed method built on computing quantum fluctuations about classical saddle solutions \cite{Zinn-Justin:1989rgp,Rubakov:2002fi,Manton:2004tk,Weinberg:2012pjx}. 
This process is most explicit in special limits, achieved by tuning parameters in the Lagrangian, for example with various levels of supersymmetry. In such limits there are dramatic simplifications in the classical equations of motion, and correspondingly also in the fluctuation problem. Beyond such special limits, semiclassical analysis is much more difficult to analyze fully. 
For example, in Yang-Mills-Higgs theory, the  classical equations of motion for monopoles and dyons 
can be reduced by a symmetric static ansatz to a set of coupled nonlinear ODEs which involves a parameter $\beta$ which is the ratio of the Higgs mass to the W-boson mass \cite{thooft,polyakov}. When $\beta\to 0$, the BPS limit, these equations reduce to first order equations with simple explicit solutions \cite{prasad,bogomolny,julia-zee}, and their fluctuations are also well understood \cite{Manton:2004tk,Weinberg:2012pjx}. This is an example of a general phenomenon \cite{bogomolny}. Away from the BPS limit, the monopole and dyons solutions are considerably more complicated, and even less is known about their fluctuations.

Here we study the 't Hooft-Polyakov monopole and dyon equations away from the BPS limit using ideas from resurgent asymptotics. 
Previous work on non-BPS monopole solutions has been numerical \cite{Bogomolny:1976ab, bais_primack:1976} and/or studying small or large deviations from the BPS limit \cite{zachos,gardner,forgacs}. In this paper we apply a method that does not rely on such approximations.
This is a general method developed by O. Costin in the context of resurgent asymptotic analysis of coupled nonlinear differential equations \cite{costin-dmj,costin-book}. The coupled nonlinear ODEs of the 't Hooft-Polyakov ansatz are naturally solved by a transseries ansatz that expands the solutions in the far field limit as a series of exponentially suppressed terms multiplied by fluctuation expansions. All higher order terms in the transseries are expressed in terms of the leading fluctuation solution. The transseries parameters are simply the coefficients of the leading homogeneous solutions, which then propagate through the transseries in a systematic fashion. This method also applies to the more general dyon solutions \cite{julia-zee}, with a more general transseries structure.

In Section II we recall some basic notation and results. Section III describes the transseries structure for non-BPS monopoles for $\beta=1$, when the Higgs and W-boson masses are equal. Section IV determines the transseries parameters by matching to a solution from small $r$. Sections V and VI describe the more general transseries structures that arise for general $\beta$, and for dyon solutions.

\section{'t Hooft-Polyakov Monopoles}
\label{sec:tp}

We consider $SU(2)$ Yang-Mills-Higgs theory with the Higgs field in the adjoint representation, for which the action is \cite{thooft,polyakov,Manton:2004tk,Rubakov:2002fi,Weinberg:2012pjx}:
\begin{equation}
    S = \int d^4 x \ \left(-\frac{1}{4e^2}F^{a}_{\mu\nu}F^{a\ \mu\nu} + \frac{1}{2}(D_{\mu}\Phi^a)(D^{\mu}\Phi^a) - \frac{\gamma}{8}(\Phi^a\Phi^a-v^2)^2 \right)
    \label{eq:action}
\end{equation}
with $a = (1,2,3)$. Here, $e$ denotes the gauge coupling constant, with $\gamma$ the strength of the scalar self-interaction and $v$ the vacuum expectation value of the Higgs field. The mass of the Higgs field is $M_H = v\sqrt{\gamma}$, with the W-bosons acquiring a mass $M_W = ve$ via the Higgs mechanism. We define the dimensionless mass ratio $\beta \equiv M_H/M_W$. The field strength tensor and covariant derivative are 
\begin{equation}
    F^a_{\mu\nu} = \partial_{\mu}A_{\nu}^a-\partial_{\nu}A_{\mu}^a+\epsilon^{abc}A^b_{\mu}A^c_{\nu}
    \qquad; \qquad D_{\mu} \Phi^a= \partial_{\mu} \Phi^a + \epsilon^{abc}A_{\mu}^b\Phi^c
\end{equation}
For monopoles we adopt the 't Hooft-Polyakov static, purely magnetic, and spherically symmetric ansatz \cite{thooft,polyakov}
\begin{equation}
    A_0^a = 0
    \qquad;\qquad
     A_i^a = \epsilon_{iak} \frac{x_k}{r^2} (W(r)-1)
    \qquad;\qquad \Phi^a = H(r) \frac{x_a}{r}
    \label{eq:ansatz}
\end{equation}
(We discuss dyon solutions in Section \ref{sec:dyons}.) In terms of the radial fields,  the monopole mass is:
\begin{equation}
    \mathcal M_\beta = \frac{4\pi v}{e} \int_0^{\infty} dr \ \left[W'^2+\frac{r^2}{2}H'^2+\frac{(1-W^2)^2}{2r^2}+\frac{\beta^2r^2}{8}(H^2-1)^2+W^2H^2 \right]
    \label{eq:mass}
\end{equation}
The classical equations of motion are
\begin{eqnarray}
    W''(r)-W(r)H(r)^2-\frac{W(r)(W(r)^2-1)}{r^2} &=& 0
    \label{eq:weq}
    \\
    H''(r) - \frac{2}{r^2}H(r)W(r)^2 - \frac{\beta^2}{2}H(r)(H(r)^2-1)+\frac{2H'(r)}{r} &=&0
    \label{eq:heq}
\end{eqnarray}
Finiteness of the mass requires the leading behavior as $r\to\infty$:
\begin{eqnarray}
    W(r) \sim e^{-r} +\dots 
    \qquad ; \qquad
    H(r) \sim 1 + \dots 
    \label{eq:large-r}
\end{eqnarray}
The precise form of the subleading terms is discussed below in Section \ref{sec:non-bps-beta1}.
In this paper we apply Costin's transseries expansion method \cite{costin-dmj,costin-book} to analyze the coupled nonlinear ODEs in \eqref{eq:weq}-\eqref{eq:heq}.

\subsection{BPS Monopole: $\beta=0$}
\label{sec:bps}

 When $\beta=0$, the monopole mass can be written in factored form
\begin{equation}
    \mathcal M_{\beta=0} = \frac{4\pi v}{e} \int_0^{\infty} dr \ \left[\left(W'+H\, W\right)^2+\frac{r^2}{2}\left(H'-\frac{1}{r^2}\left(1-W^2\right)\right)^2+\frac{d}{dr}\left(H\left(1-W^2\right)\right)\right]
    \label{eq:bps-mass}
\end{equation}
In the $\beta\to 0$ limit, the classical equations \eqref{eq:weq}-\eqref{eq:heq} reduce to first-order (BPS) equations (retaining the boundary condition that $H(r)\to 1$ at infinity) \cite{bogomolny,prasad}:
\begin{eqnarray}
W'(r) = - H(r)\, W(r) \qquad ; \qquad
H'(r) = \frac{1}{r^2}\left(1-W(r)^2\right)
\label{eq:bps-eqs}
\end{eqnarray}
This leads to the explicit closed-form BPS solutions \cite{bogomolny,prasad}:
\begin{eqnarray}
    W_{\rm BPS}(r) = \frac{r}{\sinh(r)}
    \qquad ; \qquad
    H_{\rm BPS}(r) = \coth(r)-\frac{1}{r}
    \label{eq:bps-sols}
\end{eqnarray}
As $r\to\infty$ the BPS solutions are expanded in powers of $e^{-r}$: 
\begin{eqnarray}
 W_{\rm BPS}(r) = 2r\sum_{k=0}^{\infty}e^{-(2k+1)r}
\qquad; \qquad 
H_{\rm BPS}(r) = 1-\frac{1}{r}+2\sum_{k=1}^{\infty} e^{-2kr}
\label{eq:bps-sols-infinity}
\end{eqnarray}
These are simple, convergent, transseries expansions, without asymptotic fluctuation factors.
The expansions at $r=0$ have a finite radius of convergence, $r_c=\pi$:
\begin{eqnarray}
  W_{\rm BPS}(r) =  1-\frac{r^2}{6}+\frac{7 r^4}{360}+O\left(r^6\right)
  \qquad ; \qquad
  H_{\rm BPS}(r) =\frac{r}{3}-\frac{r^3}{45}+\frac{2 r^5}{945}+O\left(r^7\right)
\end{eqnarray}
The BPS mass is determined by the field values at infinity: $W_{\rm BPS}\to 0, H_{\rm BPS}\to 1\quad \Rightarrow \quad \mathcal M_{\rm BPS}=\frac{4\pi v}{e}$.

\section{Non-BPS Monopoles: $\beta=1$}
\label{sec:non-bps-beta1}

When $\beta>0$, the finiteness of the mass requires the leading asymptotic behavior at infinity to be as in \eqref{eq:large-r}, with corrections that are exponentially small. This leads to a transseries expansion as $r\to\infty$, with sums over two independent exponential factors, $e^{-r}$ and $e^{-\beta r}$, and with power law fluctuations. To understand more clearly this resurgent transseries structure, we first specialize to the case where the Higgs and W-boson masses are equal:  $\beta=1$. This is in some sense `intermediate' between the $\beta\to 0$ BPS limit \cite{bogomolny,prasad} and the $\beta\to\infty$ limit studied in \cite{forgacs,zachos}.

\subsection{Transseries Ansatz for $\beta=1$}

When $\beta = 1$, the classical equations of motion \eqref{eq:weq}-\eqref{eq:heq} are
\begin{eqnarray}
W''&=& W\left( \frac{W^2-1}{r^2}+H^2\right)
\label{eq:Weqn}
\\
H''&=&2 H \left(\frac{1}{4} \left(H^2-1\right)+\frac{W^2}{r^2}\right) -\frac{2 H'}{r}
\label{eq:Heqn}
\end{eqnarray}
We invoke a large $r$ transseries ansatz:
\begin{eqnarray}
    W(r)&=& \sum_{k=1}^\infty e^{-k r} w_k(r)
    \label{eq:w-trans}
    \\
     H(r)&=&1- \sum_{k=1}^\infty e^{-k r} h_k(r)
     \label{eq:h-trans}
\end{eqnarray}
Each field is expanded as a sum of powers of the single exponential factor (there is only one exponential factor, because of the choice of $\beta = 1$). Each term in the exponential sum is multiplied by a ``fluctuation factor'', $w_k(r)$ or $h_k(r)$, which are generically (in fact, all except $h_1(r)$) formal factorially divergent asymptotic series as $r\to\infty$. This is very different from the BPS ``transseries'' solutions in \eqref{eq:bps-sols-infinity}, where the $w_k(r)$ and $h_k(r)$ factors truncate to single terms.
The non-BPS fluctuation factors $w_k(r)$ and $h_k(r)$ in \eqref{eq:w-trans}-\eqref{eq:h-trans} are determined recursively by the
classical equations of motion \eqref{eq:Weqn}-\eqref{eq:Heqn}.

\subsection{Leading transseries order: homogeneous linear ODEs}

At leading transseries order, $O(e^{-r})$, we obtain {\it linear and homogeneous} ODEs for $w_1(r)$ and $h_1(r)$:
\begin{eqnarray}
   w_1^{\prime\prime}(r) -2 w_1^\prime(r)+\frac{w_1(r)}{r^2}=0
   \label{eq:w1eq}
\end{eqnarray}
\begin{eqnarray}
   h_1^{\prime\prime}(r)+ \frac{2 h_1^\prime(r)}{r}-2 h_1^\prime(r)-\frac{2 h_1(r)}{r}=0
   \label{eq:h1eq}
\end{eqnarray}
The two independent real solutions for $w_1$ can be written in terms of modified Bessel functions with {\it imaginary index}:
\begin{eqnarray}
    w_{(1,\pm)}(r) = \begin{cases}
        +: \quad \sqrt{\frac{2}{\pi}}\sqrt{r}\,e^r\,\left[ I_{i\sqrt{3}/2}(r) - \frac{\sinh(\sqrt{3}\pi/2)}{i\pi}K_{i \sqrt{3}/2}(r)\right] \\ \\
        -: \quad \sqrt{\frac{2}{\pi}}\sqrt{r}\, e^r\, K_{i \sqrt{3}/2}(r)
    \end{cases}
    \label{eq:w1pm}
\end{eqnarray}
The homogeneous solution $w_{(1,-)}(r)$ decays at infinity, while $w_{(1,+)}(r)$ grows exponentially fast. So we choose 
\begin{equation}
    w_1(r) = \sigma_w\, w_{(1,-)}(r)
    \label{eq:w1sol}
\end{equation}
where $\sigma_w$ is an arbitrary constant, since \eqref{eq:w1eq} is homogeneous. At large $r$, $w_1(r)$ has a formal asymptotic expansion:
\begin{eqnarray}
    w_1(r) &\sim& \sigma_w\sum_{n=0}^{\infty} \frac{\left(\frac{1}{2}-\frac{i\sqrt{3}}{2} \right)_n\left(\frac{1}{2}+\frac{i\sqrt{3}}{2} \right)_n}{(-2)^n\,n!\,r^n}
    \qquad, \qquad r\to\infty
\nonumber\\
&\sim& \sigma_w \left[1-\frac{1}{2 r}+\frac{3}{8 r^2}-\frac{7}{16 r^3}+\frac{91}{128
   r^4}-\frac{1911}{1280 r^5}+O\left(\frac{1}{r^6}\right)\right]
    \label{eq:w1-series}
\end{eqnarray}
Here $(a)_n$ denotes Pochhammer's symbol: $(a)_n:=\Gamma(n+a)/\Gamma(a)$. The expansion \eqref{eq:w1-series} has factorially divergent coefficients with large order ($n\to \infty$) behavior:
\begin{eqnarray}
    \frac{\left(\frac{1}{2}-\frac{i\sqrt{3}}{2} \right)_n\left(\frac{1}{2}+\frac{i\sqrt{3}}{2} \right)_n}{(-2)^n\,n!}
    \sim (-1)^n&& \frac{\cosh\left(\frac{\sqrt{3}\pi}{2}\right)}{\pi (-2)^n}\Gamma(n) \notag
    \\
    &&\times\left[1+\frac{\frac{1}{2}(-2)}{(n-1)}+ \frac{\frac{3}{8}(-2)^2}{(n-1)(n-2)} + \frac{\frac{7}{16}(-2)^3}{(n-1)(n-2)(n-3)}+\cdots\right]
    \label{eq:w1n}
\end{eqnarray}
Comparing the coefficients, $\{\frac{1}{2}, \frac{3}{8}, \frac{7}{16}, \dots\}$, in \eqref{eq:w1-series} and \eqref{eq:w1n}, we see the generic resurgent large-order/low-order duality \cite{costin-book,howls,Marino:2012zq,Aniceto:2018bis,dunne-cern} between the large $n$ dependence of the expansion coefficients and the large $r$ dependence of the function in \eqref{eq:w1-series}.

The two independent  solutions of the $h_1$ equation \eqref{eq:h1eq} are much simpler:
\begin{eqnarray}
    h_{(1,\pm)}(r) = \begin{cases}
        +: \quad e^{2r}/2r \\
        -: \quad 1/r
    \end{cases}
    \label{eq:h1pm}
\end{eqnarray}
$h_{(1,-)}(r)$ decays at infinity, while $h_{(1,+)}(r)$ grows exponentially fast at infinity. So finiteness selects the solution to be
\begin{equation}
    h_1(r) = \sigma_h\, h_{(1,-)}(r)
    \label{eq:h1sol}
\end{equation}
where $\sigma_h$ is the other arbitrary constant for the $r \to \infty$ solutions. Note that $h_1(r)$  does not have a formal asymptotic series expansion as $r\to \infty$: the expansion truncates with just a single term.

\subsection{Higher transseries orders: inhomogeneous linear ODEs}

At higher transseries order, $O(e^{-k r})$ with $k\geq 2$, the transseries ansatze \eqref{eq:w-trans}-\eqref{eq:h-trans} lead to {\it linear}  {\it inhomogeneous} ODEs for $w_k(r)$ and $h_k(r)$ \cite{costin-dmj,costin-book}.

\subsubsection{Second Transseries Order}
At second transseries order, $O(e^{-2 r})$, we obtain the following linear inhomogeneous equations for $w_2(r)$ and $h_2(r)$:
\begin{eqnarray}
    w''_2(r) - 4w_2'(r) + \frac{w_2(r)}{r^2} + 3w_2(r) &=& \mathcal W_2(r)
    \label{eq:w2eq}
    \\
    h''_2(r) + \frac{2h_2'(r)}{r} - 4h_2'(r) - \frac{4h_2(r)}{r} + 3h_2(r) &=& \mathcal H_2(r)
    \label{eq:h2eq}
\end{eqnarray}
The inhomogeneous terms $\mathcal W_2(r)$ and $\mathcal H_2(r)$ mix the lower-order solutions $w_1(r)$ and $h_1(r)$: 
\begin{eqnarray}
  \mathcal W_2(r)&=&  -2h_1(r) w_1(r) 
\label{eq:ww2}
\\
\mathcal H_2(r)&=&  \frac{3}{2}h_1(r)^2+\frac{2}{r^2} w_1(r)^2
    \label{eq:hh2}
    \end{eqnarray}
Each of \eqref{eq:w2eq}-\eqref{eq:h2eq} can be solved straightforwardly using the associated homogeneous solutions
\cite{bender}. In fact, the homogeneous solutions at second order are related in a very simple way to the leading order (homogeneous) solutions:
\begin{eqnarray}
    w_{(2,\pm)}^{\rm homog}(r)=e^r w_{(1,\pm)}(r)
    \qquad, \qquad 
    h_{(2,\pm)}^{\rm homog}(r)=e^r h_{(1,\pm)}(r)
\end{eqnarray}
Furthermore, the Wronskians are simple:
\begin{eqnarray}
{\mathfrak W}\left[w_{(2,-)}^{\rm homog}(r),w_{(2,+)}^{\rm homog}(r)\right]=\frac{2}{\pi}e^{4r}
\qquad, \qquad 
{\mathfrak W}\left[h_{(2,-)}^{\rm homog}(r),h_{(2,+)}^{\rm homog}(r)\right]=\frac{e^{4r}}{r^2}
\end{eqnarray}
Therefore the second order solutions can be written explicitly in terms of the leading order solutions:
\begin{eqnarray}
    w_2(r)&=&-e^r w_{(1,+)}(r)\int_r^\infty dt\, e^{-3t} w_{(1,-)}(t)\, \mathcal W_2(t)
    +e^r w_{(1,-)}(r)\int_r^\infty dt\, e^{-3t} w_{(1,+)}(t)\, \mathcal W_2(t)
    \label{eq:w2}
    \\
    h_2(r)&=&-e^r h_{(1,+)}(r)\int_r^\infty dt\, t^2e^{-3t} h_{(1,-)}(t)\, \mathcal H_2(t)
    +e^r h_{(1,-)}(r)\int_r^\infty dt\, t^2 e^{-3t} h_{(1,+)}(t)\, \mathcal H_2(t)
    \label{eq:h2}
\end{eqnarray}
The inhomogeneous terms, $\mathcal W_2$ and $\mathcal H_2$, are also expressed in terms of the first order solutions:
\begin{eqnarray}
    \mathcal W_2(r)&=&-2\sigma_w\sigma_h h_{(1,-)}(r)w_{(1,-)}(r)
    \label{eq:www2}
    \\
    \mathcal H_2(r)&=&\frac{3}{2}\sigma_h^2 \left[h_{(1,-)}(r)\right]^2+\frac{2}{r^2}\sigma_w^2\left[w_{(1,-)}(r)\right]^2
    \label{eq:hhh2}
\end{eqnarray}
This shows that the fluctuations at second transeries order, $w_2(r)$ and $h_2(r)$ in \eqref{eq:w2}-\eqref{eq:h2}, are completely determined by the leading transeries order (homogeneous) solutions, $w_{(1,\pm)}(r)$ and $h_{(1,\pm)}(r)$ in \eqref{eq:w1pm} and \eqref{eq:h1pm}, respectively. The constants $\sigma_w$ and $\sigma_h$ enter the second order solutions via the inhomogeneous terms inside the integrals, thereby mixing the solutions beyond the leading transseries order. 

While the integrals in \eqref{eq:w2}-\eqref{eq:h2} cannot be done in closed form, it is straightforward to expand the integrands at large $t$ in order to generate the large $r$ expansions of the fluctuations $w_2(r)$ and $h_2(r)$:
\begin{eqnarray}
    w_2(r) &\sim& \sigma_w \sigma_h \left(-\frac{2}{3r} + \frac{11}{9r^2} - \frac{307}{108r^3} + \frac{1903}{216r^4}-\frac{182185}{5184 r^5} + \cdots \right)
    \label{eq:w2asymptotic}
    \\
    h_2(r)&\sim& \sigma_h^2 \left(\frac{1}{2r^2}-\frac{2}{3r^3}+\frac{13}{9r^4}-\frac{40}{9r^5}+\frac{484}{27r^6}-\frac{7280}{81r^7}+ \cdots \right) \notag \\
    && \hspace{0.5cm} +\sigma_w^2\left(\frac{2}{3r^2}-\frac{14}{9r^3}+\frac{118}{27r^4}-\frac{821}{54r^5}+\frac{5260}{81r^6}-\frac{1624441}{4860r^7} + \cdots \right)
    \label{eq:h2asymptotic}
\end{eqnarray}
The asymptotic series multiplied by $\sigma_h^2$ above in \eqref{eq:h2asymptotic} can be written in closed form, and is the sum of two incomplete gamma functions:
\begin{eqnarray}
    h_2(r) &\sim& \sigma_h^2\left( \frac{3}{4}\frac{e^r}{r} \Gamma(0,r)-\frac{3}{4} \frac{e^{3r}}{r}\Gamma(0, 3r)\right) + \sigma_w^2(\cdots)
    \label{eq:h2incompletegamma}
    \\
    &\sim& \sigma_h^2\frac{3}{4} \sum_{n=0}^{\infty}\frac{(-1)^n n!}{r^{n+2}} \left(1-\frac{1}{3^{n+1}} \right)+ \sigma_w^2(\cdots)
\end{eqnarray}
The $\sigma_h^2$ term in \eqref{eq:h2asymptotic}-\eqref{eq:h2incompletegamma} corresponds to Eq (35) in \cite{forgacs}. This is because they considered a limit in which $W(r)$ decoupled from $H(r)$. But this is only part of the full asymptotic solution.
The other terms in \eqref{eq:w2asymptotic}-\eqref{eq:h2asymptotic}, multiplying $\sigma_w \sigma_h$ and $\sigma_w^2$, do not have such simple closed-form expressions, but it is straightforward to generate their large $r$ expansions by expanding products of the modified Bessel functions $I_{i\sqrt{3}/2}(r)$ and $K_{i\sqrt{3}/2}(r)$. This general pattern continues to all higher orders: the large $r$ expansion coefficients are completely determined by the asymptotic expansions of the homogeneous solutions $w_{(1,\pm)}(r)$ and $h_{(1,\pm)}(r)$.

\subsubsection{General Transseries Order}
At higher order in the transseries, $e^{-kr}$ for $k\geq 2$, the differential equations (\ref{eq:Weqn}, \ref{eq:Heqn}) take the following general linear inhomogeneous form:
\begin{eqnarray}
    \mathcal D_k^{w} w_k(r) &=& \mathcal W_k(r) \label{eq:Wgen}\\
    \mathcal D_k^{h} h_k(r) &=& \mathcal H_k(r) \label{eq:Hgen}
\end{eqnarray}
The differential operators are:
\begin{eqnarray}
\mathcal{D}^w_k &:=& \frac{d^2}{dr^2} - 2k\frac{d}{dr} + \frac{1}{r^2} + k^2-1
\label{eq:Dw}
\\
\mathcal{D}^h_k &:=& \frac{d^2}{dr^2} + \left(\frac{2}{r} - 2k\right)\frac{d}{dr} -\frac{2k}{r} + k^2-1
\label{eq:Dh}
\end{eqnarray}
These operators have a simple relation to the leading order operators in \eqref{eq:w1eq}-\eqref{eq:h1eq}:
\begin{eqnarray}
    \mathcal{D}^w_k[e^{(k-1)r}f(r)] 
    &=& e^{(k-1)r}\mathcal{D}^w_1 f(r)
    \label{eq:dwtest}
    \\
    \mathcal{D}^h_k[e^{(k-1)r}f(r)] 
    &=& e^{(k-1)r}\mathcal{D}^h_1 f(r)
    \label{eq:dhtest}
\end{eqnarray}
This implies that the homogeneous solutions for \eqref{eq:Wgen}-\eqref{eq:Hgen} at order $k$ are directly related to the first order homogeneous solutions in \eqref{eq:w1pm} and \eqref{eq:h1pm}:
\begin{eqnarray}
    w_{(k,\pm)}^{\rm homog}(r) &=& e^{(k-1)r}\, w_{(1,\pm)}(r)\\
    h_{(k,\pm)}^{\rm homog}(r) &=& e^{(k-1)r}\, h_{(1,\pm)}(r)
\end{eqnarray}
Correspondingly, the order $k$ Wronskians
are straightforward generalizations of the  Wronskians at leading transseries order:
\begin{eqnarray}
    {\mathfrak W}^w_k(r) &:=& {\mathfrak W}\left[w^{\rm homog}_{(k,-)}(r), w^{\rm homog}_{(k,+)}(r)\right]
    =\frac{2}{\pi}e^{2kr}
    \\
    {\mathfrak W}^h_k(r) &:=& {\mathfrak W}\left[h^{\rm homog}_{(k,-)}(r), h^{\rm homog}_{(k,+)}(r)\right]=\frac{e^{2kr}}{r^2}
\end{eqnarray}
Therefore, we can write the solutions for $w_k(r)$ and $h_k(r)$ in an analogous form to those for $w_2(r)$ and $h_2(r)$ in \eqref{eq:w2}-\eqref{eq:h2}.
\begin{eqnarray}
    w_k(r) &=& -w_{(k,+)}^{{\rm homog}}(r)\int_r^\infty \frac{w_{(k,-)}^{{\rm homog}}(t) \, \mathcal W_k(t)}{{\mathfrak W}^w_k(t)}dt
    +w_{(k,-)}^{{\rm homog}}(r)\int_r^\infty \frac{w_{(k,+)}^{{\rm homog}}(t) \, \mathcal W_k(t)}{{\mathfrak W}^w_k(t)}dt \label{eq:Witerative}
    \\
    h_k(r) &=& -h_{(k,+)}^{{\rm homog}}(r)\int_r^\infty \frac{h_{(k,-)}^{{\rm homog}}(t) \, \mathcal H_k(t)}{{\mathfrak W}^h_k(t)}dt+h_{(k,-)}^{{\rm homog}}(r)\int_r^\infty \frac{h_{(k,+)}^{{\rm homog}}(t) \, \mathcal H_k(t)}{{\mathfrak W}^h_k(t)}dt \label{eq:Hiterative}
\end{eqnarray}
These expressions make it clear that at any order $k$ the solutions $w_k(t)$ and $h_k(t)$ are completely determined by the first order solutions $w_{(1,\pm)}(r)$ and $h_{(1,\pm)}(r)$ in \eqref{eq:w1pm} and \eqref{eq:h1pm}:
\begin{eqnarray}
    w_k(r) &=& -e^{(k-1) r} w_{(1,+)}(r)\int_r^\infty e^{-(k+1)t} w_{(1,-)}(t) \, \mathcal W_k(t) dt
    +e^{(k-1) r} w_{(1,-)}(r)\int_r^\infty e^{-(k+1)t} w_{(1,+)}(t) \, \mathcal W_k(t) dt \label{eq:wk}
    \\
    h_k(r) &=& -e^{(k-1) r} h_{(1,+)}(r)\int_r^\infty t^2 e^{-(k+1)t} h_{(1,-)}(t) \, \mathcal H_k(t) dt
    +e^{(k-1) r} h_{(1,-)}(r)\int_r^\infty t^2 e^{-(k+1)t} h_{(1,+)}(t) \, \mathcal H_k(t) dt \label{eq:hk}
\end{eqnarray}
Note that the inhomogeneous terms $\mathcal W_k(r)$ and $\mathcal H_k(r)$ are sums of products of the lower order solutions.
For example, the second order inhomogeneous terms are in \eqref{eq:ww2}-\eqref{eq:hh2}, while at third order we find:
\begin{eqnarray}
    \mathcal{W}_3 &=& h_1(r)^2w_1(r) + 2h_2(r)w_1(r) + \frac{1}{r^2}w_1(r)^3 - 2h_1(r)w_2(r)
    \label{eq:ww3}
    \\
    \mathcal{H}_3 &=& \frac{4}{r^2}w_1(r)w_2(r)-\frac{1}{2}h_1(r)^3-3h_1(r)h_2(r)-\frac{2}{r^2} h_1(r) w_1(r)^2
    \label{eq:hh3}
\end{eqnarray}
This demonstrates that the transseries ansatze in \eqref{eq:w-trans}-\eqref{eq:h-trans} are consistent, and that all the fluctuation terms $w_k(r)$ and $h_k(r)$, for $k\geq 2$, are determined by the leading order homogeneous solutions. 

\section{Solutions at the Origin and Matching}
\label{sec:matching}

The radial functions $W(r)$ and $H(r)$ in \eqref{eq:ansatz} have convergent Taylor series expansions about $r=0$:
\begin{eqnarray}
    W(r) &=& 1+ br^2 + \frac{r^4}{10}(a^2+3b^2) +\frac{r^6}{280}  \left(24 a^2 b-a^2+28 b^3\right)+\dots\label{eq:Wpowerseries}
    \\
    H(r) &=& ar + \frac{r^3}{20}a(8b-1)+\frac{r^5}{1120}a\left(36 a^2+192 b^2-16 b+1\right)+\dots\label{eq:Hpowerseries}
\end{eqnarray}
\begin{figure}[h!]
    \centering
    \includegraphics[width=0.5\linewidth]{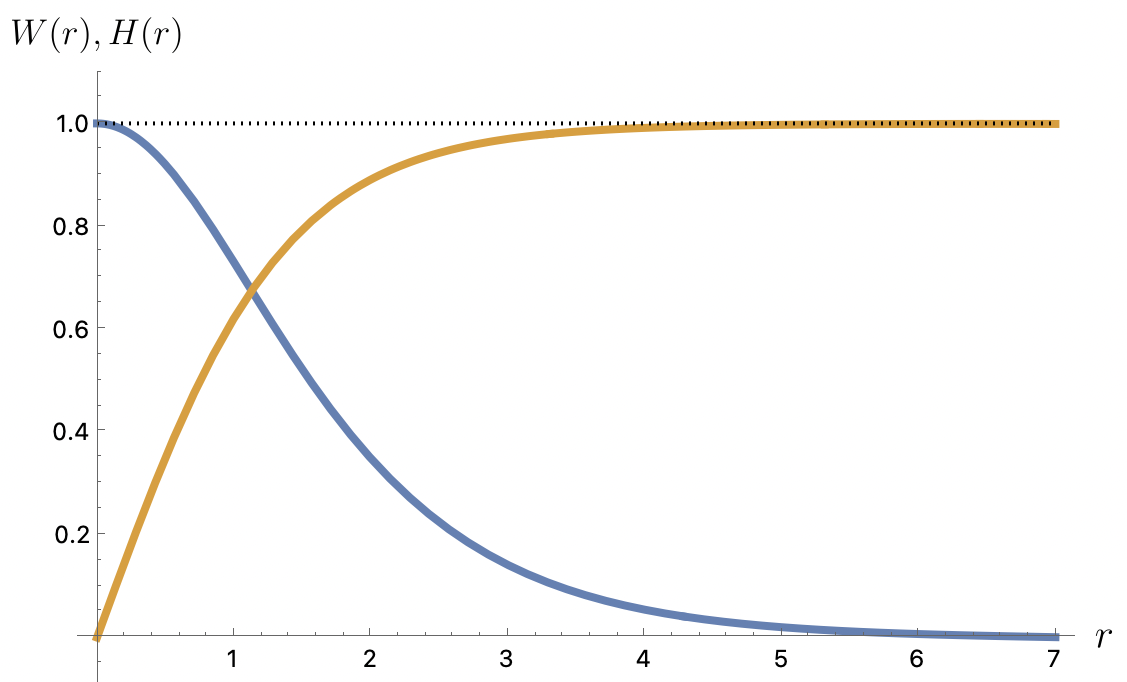}
    \caption{Numerical solutions generated from the origin. We use a combination of explicit fourth-order Runge-Kutta and linearly implicit Euler methods due to the stiffness of the differential equations \eqref{eq:Weqn}-\eqref{eq:Heqn}. $W(r)$ is given by the blue curve, and $H(r)$ by the orange curve. The asymptotic value of 1 is denoted by a dashed black line. These solutions are shown in the region $r \in (10^{-6},7).$}
    \label{fig:numerical_slns}
\end{figure}
All subsequent coefficients (and hence the radius of convergence) are determined by the two parameters, $a$ and $b$. These parameters are then determined by matching to the transseries parameters $\sigma_w$ and $\sigma_h$ in the large $r$ transseries discussed in the previous section. 
Recall that the transseries parameters 
$\sigma_w$ and $\sigma_h$ enter each order of the transseries expansions, and become mixed between the two classical equations \eqref{eq:Weqn}-\eqref{eq:Heqn} at higher orders. Therefore, we are able to refine $\sigma_w$ and $\sigma_h$ more precisely by including more terms of the transseries expansion.
This matching agrees with the numerical result, $a=0.7318$ and $b=-0.3409$, in \cite{forgacs}, for $\beta=1$,  and it determines the large $r$ transseries parameters to be: $\sigma_h=1.905$, $\sigma_w=3.336$.
Higher precision matching parameters can be generated if desired.

\begin{figure}[h!]
    \centering
    \includegraphics[width=0.5\linewidth]{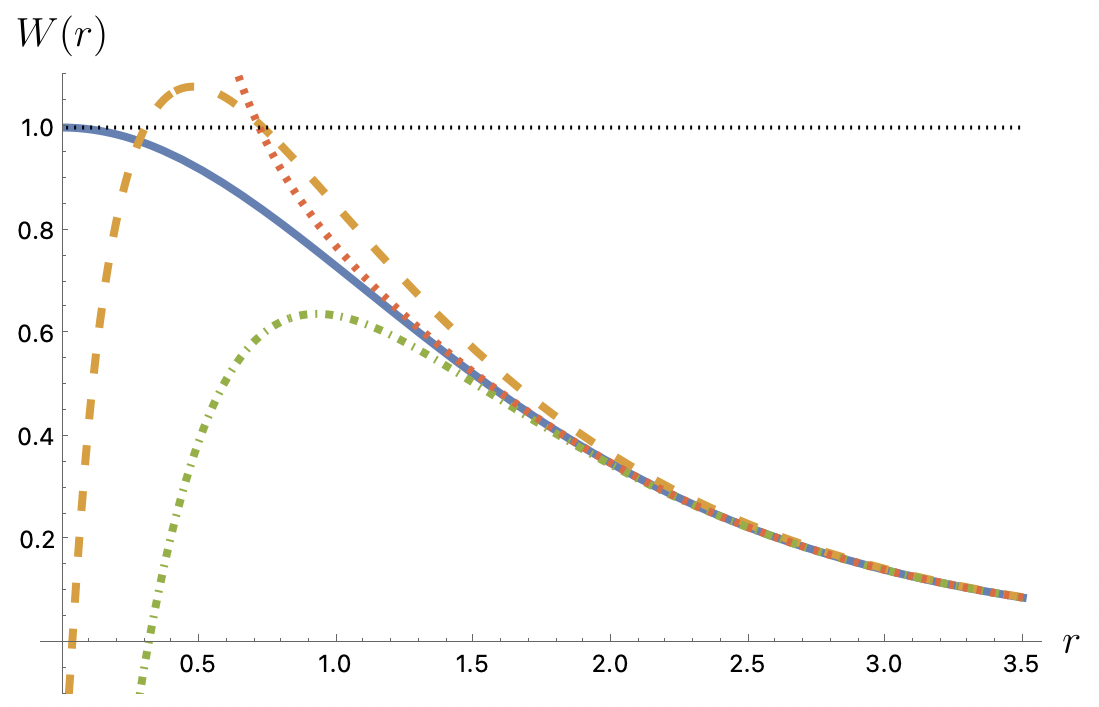}
    \caption{$W(r)$ solutions for $\sigma_h=1.905$, $\sigma_w=3.336$. The numerical solution from $r=0$ is given by the blue solid curve. First transseries order contributions are given by the yellow dashed curve, with second and third transseries order contributions given by the green dash-dotted and red dotted curves respectively. The asymptotic value of 1 is denoted by the thinly dotted black line. Note the increased agreement as $r \to 0$ with the inclusion of additional terms in the transseries.}
    \label{fig:Wsolgraph}
\end{figure}

\begin{figure}[h!]
    \centering
    \includegraphics[width=0.5\linewidth]{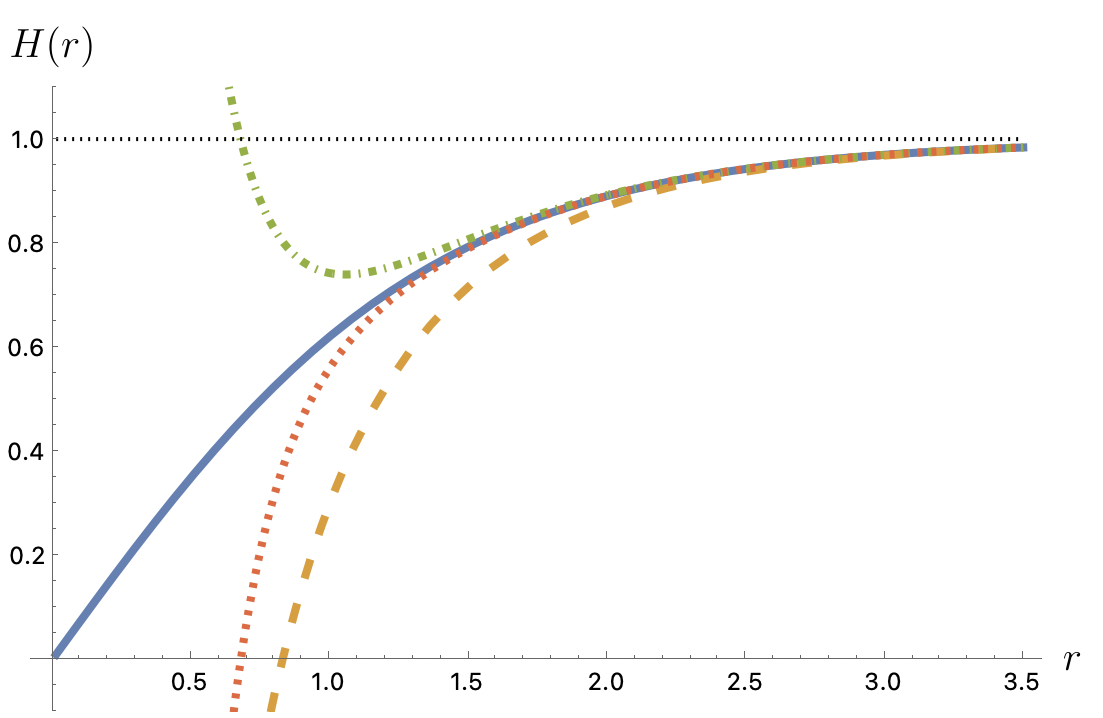}
    \caption{$H(r)$ solutions for $\sigma_h=1.905$, $\sigma_w=3.336$. The numerical solution from $r=0$ is given by the blue solid curve. First transseries order contributions are given by the yellow dashed curve, with second and third transseries order contributions given by the green dash-dotted and red dotted curves respectively. The asymptotic value of 1 is denoted by the thinly dotted black line. Note the increased agreement as $r \to 0$ with the inclusion of additional terms in the transseries.}
    \label{fig:Hsolgraph}
\end{figure}

With these values of $a$ and $b$, the 
numerical solutions for the radial profile functions $W(r)$ and $H(r)$  
are shown in Figure \ref{fig:numerical_slns}. In Figures \ref{fig:Wsolgraph} and \ref{fig:Hsolgraph} we display the effect of one, two and three terms of the transseries expansions \eqref{eq:w-trans}-\eqref{eq:h-trans}.
We observe that increasing the exponential order of the transseries expansions in \eqref{eq:w-trans}-\eqref{eq:h-trans} leads to better matching with the numerical solution at small $r$.
\begin{figure}[h!]
    \centering
    \includegraphics[width=0.5\linewidth]{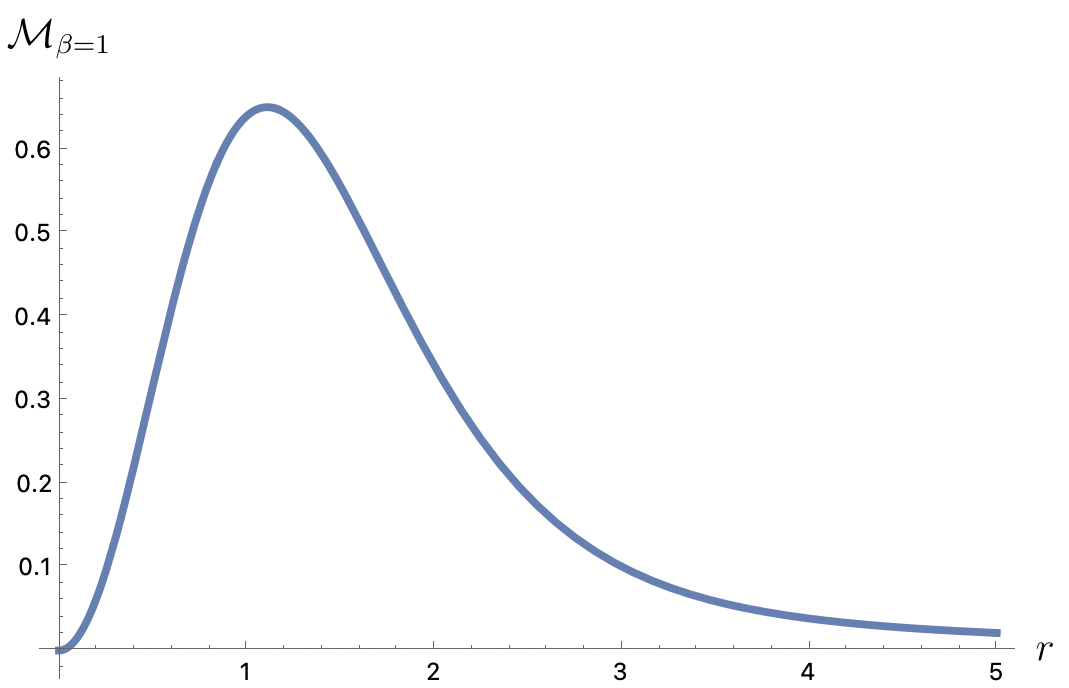}
    \caption{Integrand of the monopole mass  (\ref{eq:mass}) for $\beta = 1$. This is an intermediate curve between the $\beta \to 0$ and $\beta \to \infty$ plots presented in Figure 2 of \cite{zachos}.}
    \label{fig:integrand}
\end{figure}
Given the matched values of the parameters for the small $r$ and large $r$ expansions, we obtain 
accurate interpolating functions for $W(r)$ and $H(r)$. The resulting integrand of the monopole mass function \eqref{eq:mass} is plotted in Figure \ref{fig:integrand}. It matches well with the numerical profile plotted in \cite{zachos}.
The monopole mass can then be evaluated by integrating this interpolated solution.
We find the rescaled monopole mass for $\beta = 1$ to be
\begin{eqnarray}
    \tilde{E} = 1.237(7)
\end{eqnarray}
This is in good agreement with numerical results in \cite{Bogomolny:1976ab, forgacs}.

\section {Transseries Structure for General $\beta$}
\label{sec:beta}

Having understood the transseries structure for the situation where the Higgs and W-boson mass are equal, we now describe what changes when we consider more general values of the mass ratio, $\beta\neq 1$. 
Since $\beta$ enters the $H$ equation but not the $W$ equation, at leading transseries order only  the homogeneous $h_{(1,\pm)}$ are affected:
\begin{eqnarray}
    e^{-r} w_{(1,\pm)}(r) \rightarrow e^{-r} w_{(1,\pm)}(r)
    \qquad {\rm and}
    \qquad
    e^{-r} h_{(1,\pm)}(r) \rightarrow e^{-\beta r} h_{(1,\pm)}(\beta r)
\end{eqnarray}
Here $w_{(1,\pm)}$ and $h_{(1,\pm)}$ are the homogeneous solutions for $\beta=1$, from Section III.B. The parameter $\beta$ only enters the homogeneous $H(r)$ solutions, which are simple. The more complicated Bessel function for the homogeneous $W(r)$ solutions are not affected. 
The Wronskians are also only rescaled:
\begin{eqnarray}
   {\mathfrak W}_W(t)= \frac{2}{\pi}e^{2r}
   \qquad, \qquad 
  {\mathfrak W}_H(t)= \frac{e^{2\beta r}}{\beta r^2}
\end{eqnarray}
Therefore we can convert the coupled ODEs to integral equations, as before, and  iterate in order to generate the full transseries.

The $\beta$ dependence enters through the homogeneous $H_\pm(r)$ solutions, and this $\beta$ dependence propagates as one iterates the integral equations, at higher orders also due to the nonlinear coupling. 
Since the homogeneous $H_\pm(r)$ solutions have exponential factors $e^{\pm \beta r}$ involving $\beta$ in the exponent, while the homogeneous $W_\pm(r)$ solutions have exponential factors $e^{\pm r}$ without $\beta$ in the exponent, the iteration automatically generates a {\it double series} of exponentials involving all powers of $e^{\pm r}$ and all powers of $e^{\pm \beta r}$. 
There are still just two transseries parameters $\sigma_w$ and $\sigma_h$, but their values when matching to the expansions at the origin become $\beta$ dependent.

\section{Dyons}
\label{sec:dyons}

In addition to monopole solutions, the Yang-Mills-Higgs theory \eqref{eq:action} supports dyon solutions, which carry both electric and magnetic charge \cite{julia-zee,witten}. 
For dyons the radial ansatz is generalized to include the $A_0$ field:\footnote{Note that our radial profile function of the Higgs field, $H(r)$, differs from that in \cite{julia-zee} by a factor of $r$.}
\begin{equation}
    A_0^a = J(r)\frac{x_a}{r^2}
    \qquad;\qquad
     A_i^a = \epsilon_{iak} \frac{x_k}{r^2} (W(r)-1)
    \qquad;\qquad \Phi^a = H(r) \frac{x_a}{r}
    \label{eq:dyonansatz}
\end{equation}
We now have a system of three coupled nonlinear ODEs:
\begin{eqnarray}
    J'' &=& 2JW^2 
    \label{eq:j-dyon}
    \\
    W''&=& W\left( \frac{W^2-J^2-1}{r^2}+H^2\right)
    \label{eq:w-dyon}
    \\
    H''&=&2 H \left(\frac{\beta^2}{4} \left(W^2-1\right)+\frac{W^2}{r^2}\right) -\frac{2 H'}{r}
    \label{eq:h-dyon}
\end{eqnarray}
For simplicity of exposition, and to compare directly with the monopole transseries structure in Section \ref{sec:non-bps-beta1},  we consider again the $\beta=1$ case; noting that the strategy for general $\beta$ is as described in section \ref{sec:beta}.

The coupled equations \eqref{eq:j-dyon}-\eqref{eq:h-dyon} (with $\beta=1$) are solved at large $r$ by a transseries ansatz analogous to \eqref{eq:w-trans}-\eqref{eq:h-trans}. The zeroth exponential order is slightly different, yielding:
\begin{equation}
    j_0(r)=  \mu + \kappa r \qquad ; \qquad w_0(r) = 0 \qquad ; \qquad h_0(r) = 1
\end{equation}
Here $\mu$ and $\kappa$ are two undetermined parameters corresponding to the electric charge and mass scale of $J$, respectively. The linear behavior of $J(r)$ arises from \eqref{eq:j-dyon} because $W(r)\sim e^{-r}$, so $J''\sim 0$ at this order.

The introduction of the new mass parameter $\kappa$, which is restricted by the boundary conditions to the range $0 \leq \kappa < 1$, generates an additional exponential term $e^{-r\sqrt{1-\kappa^2}}$. This leads to a more general transseries structure:
\begin{eqnarray}
    J &=& \mu + \kappa r  + \sum_{k = 1}^{\infty}e^{-kr} j^{(1)}_k(r) + \sum_{\ell = 1}^{\infty} e^{-\ell r\sqrt{1-\kappa^2}} j^{(2)}_{\ell}(r)
    \label{eq:j-trans-dyon}
    \\
    W &=& \sum_{k = 1}^{\infty}e^{-kr} w^{(1)}_k(r) + \sum_{\ell = 1}^{\infty} e^{-\ell r\sqrt{1-\kappa^2}} w^{(2)}_{\ell}(r)
    \label{eq:w-trans-dyon}
    \\
    H &=& 1 + \sum_{k = 1}^{\infty}e^{-kr} h^{(1)}_k(r) + \sum_{\ell = 1}^{\infty} e^{-\ell r\sqrt{1-\kappa^2}} h^{(2)}_{\ell}(r)
    \label{eq:h-trans-dyon}
\end{eqnarray}
At leading $(k = \ell = 1)$ order, we find:
\begin{eqnarray}
     J &=& \mu + \kappa r  + \mathcal{O}(e^{-2r}) + \mathcal{O}(e^{-2r\sqrt{1-\kappa^2}})
    \label{eq:j-trans-dyon-simp}
    \\
    W &=& e^{-r\sqrt{1-\kappa^2}} w_{1}(r) + \mathcal{O}(e^{-2r}) + \mathcal{O}(e^{-2r\sqrt{1-\kappa^2}})
    \label{eq:w-trans-dyon-simp}
    \\
    H &=& 1 + e^{-r}h_1(r) + \mathcal{O}(e^{-2r}) + \mathcal{O}(e^{-2r\sqrt{1-\kappa^2}})
    \label{eq:h-trans-dyon-simp}
\end{eqnarray}
Since $J(r)$ does not enter the $H$ equation \eqref{eq:h-dyon}, at leading transseries order the homogeneous solution $h_1(r)$ is the same as in the monopole case. However, the differential equation for $w_1(r)$ is modified by the $j_0(r)$ term:
\begin{equation}
    w_1''(r) - 2w_1'(r) + \frac{1+(\mu+\kappa r)^2}{r^2}w_1(r) = 0
\end{equation}
The solutions are linear combinations of Whittaker functions (confluent hypergeometric functions)
\begin{equation}
    w_1(r) = c_1e^r M_{\alpha,\delta}(2r\sqrt{1-\kappa^2}) + c_2e^rW_{\alpha,\delta}(2r\sqrt{1-\kappa^2})
\end{equation}
where $\alpha ={\kappa\mu}/{\sqrt{1-\kappa^2}}$, and $\delta ={i\sqrt{3+4\mu^2}}/{2}$. In the monopole limit, $\kappa, \mu\to 0$, the Whittaker functions reduce to modified Bessel functions.
As before, at higher transseries order, $k, \ell \geq 2$, we generate {\it inhomogeneous} linear differential equations for the fluctuation factors, and these can therefore be solved iteratively in the same manner as for the monopole case.

\section{Conclusion}

We have shown that non-BPS monopole and dyon solutions to the classical equations of motion for the $SU(2)$ 't Hooft-Polyakov have a natural construction in the far field regime as resurgent transseries. Increasing the exponential order of the transseries extends the precision into the near field regime. This method also applies to other well-known sets of coupled nonlinear equations arising in quantum field theory, away from special parametric limits, such as  general kink, vortex, false vacuum decay, and instanton equations, as well as in nonlinear equations in general relativity. An immediate consequence of our result is that the zero-mode solutions, arising for fermions in background monopole fields \cite{Jackiw:1975fn}, also have resurgent transseries expansions, involving exponentials of transseries. 
Current and future work will address the connection problem, relating the far field region to the near field region, as well as the general fluctuation problem. It would also be interesting to study the transseries structure of associated soliton fields \cite{manton}, and quantum corrections to the monopole mass \cite{Kiselev:1988gf,Rajantie:2005hi}.
\bigskip

\noindent{Note added:} A few days before final completion of this paper, a paper appeared on the arXiv concerning some other resurgent features of the monopole equations, based on the large $\beta$ limit \cite{Malinsky:2026eux}.

\acknowledgements

This material is based upon work supported by the U.S. Department of Energy, Office of Science, Office of High Energy Physics under Award Number DE-SC0010339. 

\bibliography{biblio}

\end{document}